\documentclass[12pt]{article}
\usepackage{mathptmx}
\usepackage[utf8]{inputenc}
\usepackage[letterpaper, portrait, margin=1in]{geometry}
\usepackage[english]{babel}
\usepackage{amsmath,amssymb}
\usepackage{graphicx}
\usepackage{enumitem}
\usepackage{ragged2e}
\usepackage[range-units=brackets, range-phrase=-]{siunitx}
\usepackage[symbol]{footmisc}
\usepackage{multicol}
\usepackage{cite}
\usepackage{xcolor}
\definecolor{dullpurple}{rgb}{0.431,0.188,0.534}
\definecolor{darkgreen}{rgb}{0.075,0.302,0.047}
\definecolor{dullred}{rgb}{0.706,0.208,0.192}
\usepackage[colorlinks=true, urlcolor=dullpurple, citecolor=dullred, linkcolor=darkgreen]{hyperref}
\usepackage[small]{caption}
\usepackage[babel=true]{microtype}

\def\Neff{N_\mathrm{eff}}
\def\Mpl{M_\mathrm{P}}
\newcommand{\planck}{\textsl{Planck}}

\makeatletter
\renewcommand\section{\@startsection{section}{1}{\z@}%
	{-2.4ex \@plus -0.7ex \@minus -.4ex}%
	{1.8ex \@plus.4ex \@minus .4ex}%
	{\normalfont\large\bfseries}}
\renewcommand\subsection{\@startsection{subsection}{2}{\z@}%
	{-2.0ex\@plus -1ex \@minus -.5ex}%
	{1.0ex \@plus .3ex \@minus .3ex}%
	{\normalfont\normalsize\bfseries}}
\makeatother

\begin{document}

\pagenumbering{roman}
\thispagestyle{empty}
{\raggedright
\huge
Astro2020 Science White Paper \linebreak

Messengers from the Early Universe:\\[8pt]Cosmic Neutrinos and Other Light Relics
\linebreak
\normalsize

\noindent \textbf{Thematic Area:}
Cosmology and Fundamental Physics \linebreak

\textbf{Principal Author:}

Name: Daniel Green
 \linebreak						
Institution: University of California San Diego
 \linebreak
Email: \href{mailto:drgreen@physics.ucsd.edu}{drgreen@physics.ucsd.edu}
 \linebreak 
 
}

\noindent
\textbf{Abstract:} The hot dense environment of the early universe is known to have produced large numbers of baryons, photons, and neutrinos. These extreme conditions may have also produced other long-lived species, including new light particles (such as axions or sterile neutrinos) or gravitational waves. The gravitational effects of any such light relics can be observed through their unique imprint in the cosmic microwave background (CMB), the large-scale structure, and the primordial light element abundances, and are important in determining the initial conditions of the universe. We argue that future cosmological observations, in particular improved maps of the~CMB on small angular scales, can be orders of magnitude more sensitive for probing the thermal history of the early universe than current experiments. These observations offer a unique and broad discovery space for new physics in the dark sector and beyond, even when its effects would not be visible in terrestrial experiments or in astrophysical environments. A~detection of an excess light relic abundance would be a clear indication of new physics and would provide the first direct information about the universe between the times of reheating and neutrino decoupling one second later.

\clearpage
{\raggedright
\noindent\textbf{Co-authors\footnote{The list of affiliations can be found in the Appendix.}:}
Mustafa A. Amin$^{1}$, 
Joel Meyers$^{2}$, 
Benjamin Wallisch$^{3,4}$, 
Kevork N.\ Abazajian$^{5}$, 
Muntazir Abidi$^{6}$, 
Peter Adshead$^{7}$, 
Zeeshan Ahmed$^{8}$, 
Behzad Ansarinejad$^{9}$, 
Robert Armstrong$^{10}$, 
Carlo Baccigalupi$^{11,12,13}$
% ...

}

\clearpage
\pagenumbering{arabic}
\setcounter{page}{1}

\section{Introduction}

Cosmology unites the study of the fundamental laws of particle physics, the history of the universe, the origin of its structure, and its subsequent dynamics. The abundances of baryons, photons, neutrinos, and (possibly) dark matter were determined during the hot thermal phase that dominated the early universe. It is the abundances of these particles and the forces between them that determine the conditions of the cosmos that we see today. 

There is strong motivation to determine if other forms of radiation (i.e.~relativistic species), including gravitational waves, were produced during the hot big bang. Changes to the radiation density make a measurable impact on cosmological observables, including the amplitude of clustering, the scale of the baryon acoustic oscillations~(BAOs), and primordial light element abundances. An accurate measurement of the total radiation density is therefore also crucial in order to calibrate late-time observables, such as the BAO~scale or the lensing amplitude.

\emph{New sources of (dark) radiation are well motivated by both particle physics and cosmology} (cf.\ e.g.~\cite{Essig:2013lka, Marsh:2015xka, Alexander:2016aln}). New light particles are predicted in many extensions of the Standard Model~(SM), including axions and sterile neutrinos, or can arise as a consequence of solving the hierarchy problem (see e.g.~\cite{Abazajian:2001nj, Strumia:2006db, Ackerman:2008gi, Boyarsky:2009ix, Arvanitaki:2009fg, Cadamuro:2010cz, Kaplan:2011yj, Abazajian:2012ys, CyrRacine:2012fz, Brust:2013xpv, Weinberg:2013kea, Salvio:2013iaa, Essig:2013lka, Kawasaki:2015ofa, Graham:2015cka, Marsh:2015xka, Baumann:2016wac, Alexander:2016aln, Arkani-Hamed:2016rle, Chacko:2016hvu, Craig:2016lyx, Chacko:2018vss}). For large regions of unexplored parameter space, these light particles are thermalized in the early universe and lead to additional radiation at later times. Light species are ubiquitous in models of the late universe as well: they may form the dark matter (e.g.~axions), be an essential ingredient of a more complicated dark sector as the force carrier between dark matter and the Standard Model (or itself), or provide a source of dark radiation for a dark thermal history. Furthermore, these new particles could also play a role in explaining discrepancies in the measurements of the Hubble constant~$H_0$~\cite{Archidiacono:2013fha, Bernal:2016gxb, Zhang:2017aqn, Addison:2017fdm, Aylor:2018drw}, the amplitude of large-scale matter fluctuations~$\sigma_8$~\cite{MacCrann:2014wfa, Lesgourgues:2015wza, Kohlinger:2017sxk, Joudaki:2017zdt}, and the properties of clustering on small scales~\cite{Weinberg:2013aya,Aghanim:2018eyx}. Measuring the total radiation density is a broad window into all these possibilities as well as additional scenarios that we have yet to consider.

Remarkably, cosmological observations provide an increasingly sharp view of the radiation content of the universe. The cosmic neutrino background itself is a compelling example: while it has not been possible to see cosmic neutrinos in the lab, their presence has been observed at high significance in the cosmic microwave background~(CMB) and through observations of light element abundances~\cite{Aghanim:2018eyx,Cyburt:2015mya}. These indirect measurements of the cosmic neutrino background therefore provide a window back to a few seconds after the big bang, the era of neutrino decoupling. A new thermalized light particle adds at least a percent-level correction to the radiation density that is determined by its decoupling temperature (time). Measurements in the coming decade will be sensitive to decoupling temperatures that are orders of magnitude higher than current experiments, and able to \emph{reveal new physics that will be inaccessible in any other setting.}

\section{Light Relics of the Big Bang}

\subsection*{Cosmic Neutrino Background}
The cosmic neutrino background is one of the remarkable predictions of the hot big bang. In the very early universe, neutrinos were kept in thermal equilibrium with the Standard Model plasma. As the universe cooled, neutrinos decoupled from the plasma. A short time later, the relative number density and temperature in photons increased, due primarily to the transfer of entropy from electron-positron pairs to photons. The background of cosmic neutrinos persists today, with a temperature and number density similar to that of the CMB. Their energy density~$\rho_\nu$ is most commonly expressed in terms of the effective number of neutrino species,\vspace{-7pt}
\begin{equation}
\Neff = \frac{8}{7}\left(\frac{11}{4}\right)^{\!4/3} \frac{\rho_\nu}{\rho_\gamma} \, ,\vspace{-5pt}
\end{equation}
where~$\rho_\gamma$ is the energy density in photons. This definition is chosen so that $\Neff=3$ in the~SM if neutrinos had decoupled instantaneously prior to electron-positron annihilation. The neutrino density~$\rho_\nu$ receives a number of corrections from this simple picture of decoupling, and the best available calculations give $\Neff^\mathrm{SM} = 3.045$ in the~SM~\cite{Mangano:2005cc, Grohs:2015tfy, deSalas:2016ztq}.

Cosmology is sensitive to the gravitational effects of neutrinos, both through their mean energy density~\cite{Peebles:1966zz, Dicus:1982bz, Hou:2011ec, Bashinsky:2003tk} and their fluctuations, which propagate at the speed of light in the early universe due to the free-streaming nature of neutrinos~\cite{Bashinsky:2003tk, Baumann:2015rya, Baumann:2017lmt}. A radiation fluid whose fluctuations do not exceed the sound speed of the plasma~\cite{Bell:2005dr, Friedland:2007vv} could arise from large neutrino self-interactions~\cite{Cyr-Racine:2013jua, Lancaster:2017ksf}, neutrino-dark sector interactions, or dark radiation self-coupling. Such a radiation fluid can be observationally distinguished from free-streaming radiation, and can serve as both a foil for the cosmic neutrino background and a test of new physics in the neutrino and dark sectors~\cite{Baumann:2015rya, Brust:2017nmv, Choi:2018gho}.

\emph{Neutrinos are messengers from a few seconds after the big bang and provide a new window~into our cosmological history.} While these relics have been detected in cosmological data, higher precision measurements would advance the use of neutrinos as a cosmological probe. Furthermore, the robust measurement of the neutrino abundance from the CMB is crucial for inferring cosmic parameters, including the expansion history using BAOs~\cite{Dodelson:2016wal}, the neutrino masses~\cite{Abazajian:2013oma}, and~$H_0$~\cite{Aylor:2018drw}.

\subsection*{Beyond the Standard Model}

\emph{A measurement of the value of~$\Neff$ provides vastly more information than just the energy density in cosmic neutrinos.} The parameter~$\Neff$ is a probe of any particles that have the same gravitational influence as relativistic neutrinos, which is true of any (free-streaming) radiation. Furthermore, this radiation could have been created at much earlier times when the energy densities were even higher than in the cores of stars or supernovae, shedding light on the physics at new extremes of temperatures as well as densities, and our early cosmic history.

New light particles that were thermally produced in the early universe contribute to the neutrino density~$\rho_\nu$ and increase~$\Neff$ above the amount from neutrinos alone. The presence of any additional species can therefore be characterized by $\Delta\Neff \equiv \Neff - \Neff^\mathrm{SM}$. Since all such thermalized particles behave in the same way from a cosmological point of view, this parametrization captures a vast range of new physics: axions, sterile neutrinos, dark sectors, and beyond~\cite{Brust:2013xpv, Chacko:2015noa, Baumann:2016wac, Abazajian:2016yjj}.

\begin{figure}
\centering
\includegraphics[trim= 0 0 0 5]{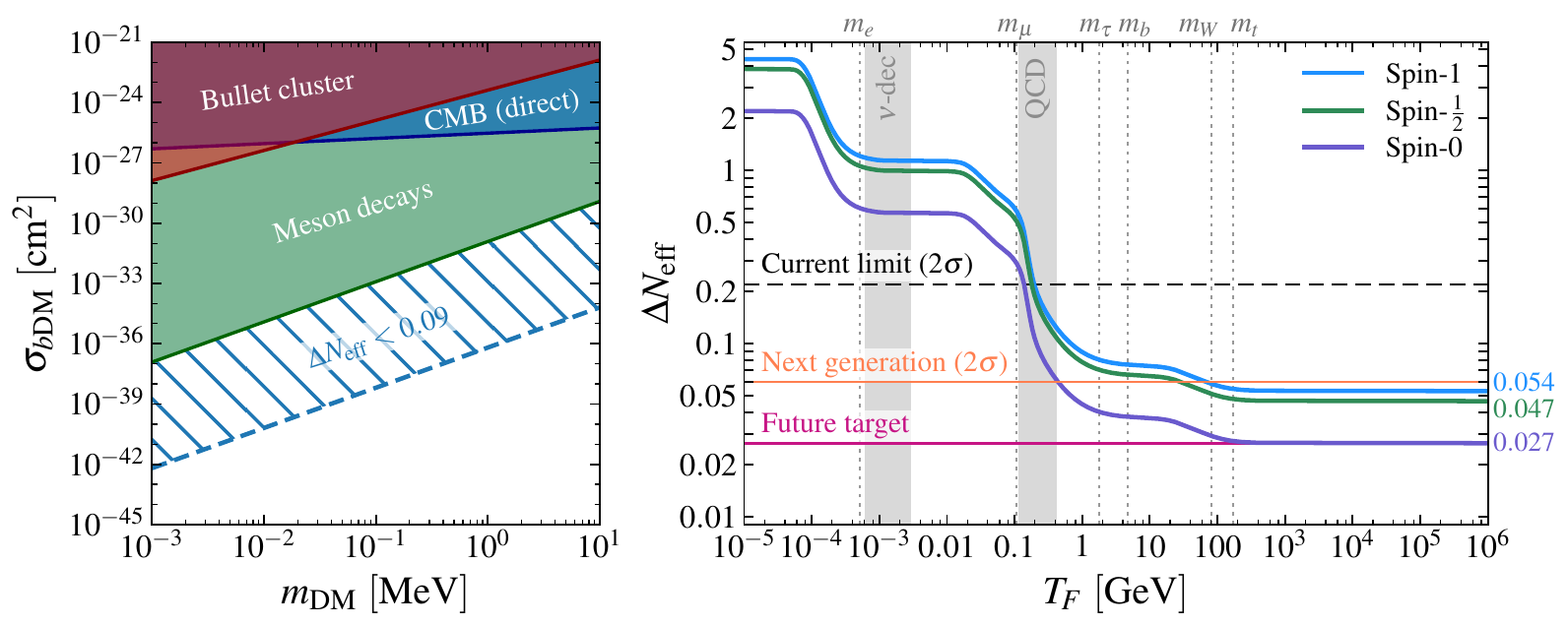}\vspace{-6pt}
\caption{\textit{Left:}~Limits on the dark matter-baryon cross section~$\sigma_{b\mathrm{DM}}$ for a Yukawa potential. Future cosmological constraints will restrict $\Delta \Neff < 0.09$ and, therefore, exclude cross sections large enough to thermalize the (\SI{200}{keV}-mass) particle mediating the force~\cite{Green:2017ybv}. This limit is compared to the direct bound on baryon-dark matter scattering from the CMB~\cite{Gluscevic:2017ywp} and to the constraints on dark forces from~the Bullet Cluster~\cite{Markevitch:2003at}. The strongest current constraint is from the absence of meson decays to the mediator~\cite{Essig:2010gu}. \textit{Right:}~Contributions of a single massless particle, which decoupled at the temperature~$T_F$ from the Standard Model, to the effective number of relativistic species, $\Neff = N_\mathrm{eff}^\mathrm{SM} + \Delta\Neff$, with the Standard Model expectation $N_\mathrm{eff}^\mathrm{SM} = 3.045$ from neutrinos. The limit at 95\%~c.l.\ from a combination of current CMB, BAO and BBN~observations~\cite{Aghanim:2018eyx}, and the anticipated sensitivity of next-generation CMB experiments (cf.~e.g.~\cite{Abazajian:2016yjj, Baumann:2017gkg, Hanany:2019lle}) illustrate the current and future power of cosmological surveys to constrain light thermal relics. The displayed values on the right are the observational thresholds for particles with different spins and arbitrarily large decoupling temperature.}\vspace{-9pt}
\label{fig:deltaNeff}
\end{figure}

Constraints on~$\Neff$ are broadly useful and, most importantly, allow the exploration of new and interesting territory in a variety of well-motivated models. This can be seen with a simple example: dark matter-baryon scattering. For low-mass (sub-\si{GeV}) dark matter, current data allows for relatively large scattering cross sections~\cite{Battaglieri:2017aum}. If they scatter through a Yukawa potential, which is a force mediated by a scalar particle, this force is consistent with fifth-force experiments and stellar cooling if the mediator has a mass around~\SI{200}{keV}. However, the particle which mediates the force necessarily\footnote{The mediator with a mass of~\SI{200}{keV} is too heavy to contribute to~$\Neff$, but it must decay to sub-\si{eV} mass particles, which will increase~$\Neff$, in order to avoid more stringent constraints.} contributes $\Delta \Neff \geq 0.09$ when it comes into thermal equilibrium with the Standard Model~\cite{Green:2017ybv}. Excluding this value would require that the strength of the interactions is small enough to prevent the particle from reaching equilibrium at any point in the history of the universe, which, consequently, limits the scattering cross section, as shown in the left panel of Fig.~\ref{fig:deltaNeff}. This measurement is sensitive to 10--15 orders of magnitude in cross section that are not probed by direct constraints from cosmology and astrophysics, and five orders of magnitude stronger than meson decay searches. We see that \emph{cosmological measurements of~$\Delta \Neff$ are an extremely sensitive probe of dark sector physics that are complementary to more direct tests, both in the laboratory and with astrophysical observations}~\cite{Green:2017ybv, Knapen:2017xzo}.

More generally, the contribution to~$\Neff$ from any thermalized new particle is easy to predict because its energy density in equilibrium is fixed by the temperature and the number of internal states (e.g.~spin configurations). Under mild assumptions (see e.g.~\cite{Wallisch:2018rzj} for a detailed discussion), the contribution to~$\Delta\Neff$ is determined by two numbers, the last temperature at which it was in equilibrium,~$T_F$, and the effective number of spin degrees of freedom,~$g_s$, according to\vspace{-8pt}
\begin{equation}
\Delta\Neff = g_s \left(\frac{43/4}{g_{\star}(T_F)}\right)^{\!4/3} .\vspace{-7pt}
\end{equation}
The function~$g_{\star}(T_F)$ is the number of effective degrees of freedom (defined as the number of independent states with an additional factor of~$7/8$ for fermions) of the SM~particle content at the temperature~$T_F$. This function appears in the formula for~$\Delta\Neff$ because it determines how much the photons are heated relative to a new light particle due to the annihilation of the heavy SM~particles as the universe cooled (see the right panel of Fig.~\ref{fig:deltaNeff}). The next generation of (proposed) CMB~observations are expected to reach a precision of $\sigma(\Neff) = 0.03$, which would extend our reach in~$T_F$ by several orders of magnitude for a particle with spin~$s > 0$ and be the first measurement sensitive to a real scalar ($s=0$) that decouples prior to the QCD~phase transition.

To understand the impact of such a measurement, recall that equilibrium at temperature~$T$ arises when the production rate~$\Gamma$ is much larger than the expansion rate~$H(T)$. At high temperatures, production is usually fixed by dimensional analysis, $\Gamma \propto \lambda^2 T^{2n+1}$, where~$\lambda$ is the coupling to the Standard Model with units of~$[\mathrm{Energy}]^{-n}$. The particle is therefore in equilibrium if $\lambda^2 \gg \Mpl^{-1} T^{-2n+1}$. There are two important features of this formula: (i)~the appearance of the Planck scale~$\Mpl$ implies we are sensitive to very weak couplings ($\Mpl^{-2} = 8 \pi G_{\hskip-1ptN}$), and (ii)~for $n \geq 1$ it scales like an inverse power of~$T$. As a result, sensitivity to increasingly large~$T_F$ implies that we are probing increasingly weak couplings (lower production rates) in proportion to the improvement in~$T_F$ (not~$\Delta\Neff$). These two features explain why future measurements of~$\Delta\Neff$ can be orders of magnitude more sensitive than terrestrial and astrophysical probes of the same physics~\cite{Baumann:2016wac, Abazajian:2016yjj}.

The impact of the coming generation of observations is illustrated in Fig.~\ref{fig:deltaNeff}. Anticipated~improvement in measurements of~$\Neff$ translate into orders of magnitude in sensitivity to the temperature~$T_F$. This temperature sets the reach in probing fundamental physics. Even in the absence of a detection, future cosmological probes would place constraints that can be orders of magnitude stronger than current probes of the same physics, including for axion-like particles~\cite{Baumann:2016wac} and dark sectors~\cite{Adshead:2016xxj, Craig:2016lyx, Green:2017ybv, Chacko:2018vss}. It is also worth noting that these contributions to~$\Neff$ asymptote to specific values of $\Delta\Neff = 0.027, 0.047, 0.054$ for a massless (real) spin-0 scalar, spin-1/2 (Weyl) fermion and spin-1 vector boson, respectively (see Fig.~\ref{fig:deltaNeff}). A cosmological probe with sensitivity to~$\Delta\Neff$ at these levels would \emph{probe physics back to the time of reheating for even a single additional species.}

Even without new light particles, $\Neff$ is a \emph{probe of new physics that changes our thermal history, including processes that result in a stochastic background of gravitational waves}~\cite{Boyle:2007zx, Stewart:2007fu, Meerburg:2015zua}. Violent phase transitions and other nonlinear dynamics in the primordial universe could produce such a background, peaked at frequencies much larger than those accessible to B-mode polarization measurements of the~CMB or, in many cases, direct detection experiments such as~LIGO and~LISA~\cite{Maggiore:1999vm, Easther:2006gt, Dufaux:2007pt, Amin:2014eta,Caprini:2018mtu}. For particularly violent sources, the energy density in gravitational waves can be large enough to make a measurable contribution to~$\Neff$~\cite{Caprini:2018mtu, Adshead:2018doq, Amin:2019qrx}.

In addition to precise constraints on~$\Neff$, cosmological probes will provide an \emph{independent high-precision measurement of the primordial helium abundance~$Y_\mathrm{p}$} due to the impact of helium on the free electron density prior to recombination. This is particularly useful since~$Y_\mathrm{p}$ is sensitive to~$\Neff$ a few minutes after the big bang, while the~CMB and matter power spectra are affected by~$\Neff$ prior to recombination, about \num{370000}~years later. Measuring the radiation content at these well-separated times provides a window onto any nontrivial evolution in the energy density of radiation in the early universe~\cite{Fischler:2010xz, Hasenkamp:2011em, Hooper:2011aj, Hasenkamp:2012ii}. Furthermore, $\Neff$ and~$Y_\mathrm{p}$ are sensitive to neutrinos and physics beyond the Standard Model in related, but different ways, which allows for even finer probes of new physics, especially in the neutrino and dark sectors.

\section{Cosmological and Astrophysical Observables}

\noindent\textbf{Cosmic Microwave Background}\hskip8pt
The effect of the radiation density on the damping tail of the anisotropy power spectrum drives the constraint on~$\Neff$ from the~CMB. The largest effect comes from the change to the expansion rate, which impacts the amount of photon diffusion, which in turn causes an exponential suppression of short wavelength modes~\cite{Zaldarriaga:1995gi}. This effect on the damping tail is dominant when holding fixed the scale of matter-radiation equality and the location of the first acoustic peak~\cite{Hou:2011ec}, both of which are precisely measured. At the noise level and resolution of upcoming observations~\cite{Benson:2014qhw, Henderson:2015nzj, Abazajian:2016yjj, Ade:2018sbj, Hanany:2019lle}, this effect is predominately measured through the TE~power spectrum on small scales. \planck\ has provided a strong constraint of $\Neff = 2.92^{+0.18}_{-0.19}$ using temperature and polarization data~\cite{Aghanim:2018eyx}. \emph{Future high-resolution maps of the CMB could realistically achieve $\mathit{\sigma(\Neff)= 0.03}$ in the coming decade}~\cite{Abazajian:2016yjj, Hanany:2019lle}.

In addition to the effect on the expansion rate, perturbations in neutrinos (and other free-streaming light relics) affect the photon-baryon fluid through their gravitational influence. The contributions from neutrino fluctuations are well described by a correction to the amplitude and the phase of the acoustic peaks in both temperature and polarization~\cite{Bashinsky:2003tk}. The phase shift is a particularly compelling signature since it is not degenerate with other cosmological parameters (unlike the damping tail)~\cite{Bashinsky:2003tk, Baumann:2015rya} and has a direct connection to the underlying particle properties~\cite{Baumann:2015rya}. Recently, the phase shift from neutrinos has also been established directly in the \planck\ temperature data~\cite{Follin:2015hya}, which provides the most direct evidence for free-streaming radiation consistent with the cosmic neutrino background. \emph{If $\Delta \Neff \neq 0$ is detected, this phase could provide a powerful confirmation.}

\vskip4pt
\noindent\textbf{Big Bang Nucleosynthesis~(BBN)}\hskip8pt
The production of light elements in the early universe is affected by the density of light relics through their impact on the expansion rate during the first~few minutes after reheating. Cosmic neutrinos play a special role during~BBN since they also participate in the weak interactions that interconvert protons and neutrons. Measurements of the primordial abundances of light elements can therefore be used to infer the relic density of neutrinos and other light species, with deuterium~\cite{Cooke:2017cwo} and \mbox{helium-4}~\cite{Aver:2015iza, Peimbert:2016bdg} currently providing the tightest constraints. Future improvements will be driven by \SI{30}{\meter}-class telescopes, but are limited by the analysis of the most pristine astrophysical systems rather than statistics. When abundance measurements are combined with \planck~CMB~data, the density of light relics is found to be $\Neff = 3.04 \pm 0.11$~\cite{Aghanim:2018eyx}.

\vskip4pt
\noindent\textbf{Large-Scale Structure~(LSS)}\hskip8pt
\emph{Maps of the large-scale structure of the universe from galaxy and weak lensing surveys can provide complementary measurements of the radiation content.} The main observable is the shape of the matter power spectrum, which can be decomposed into a smooth (broadband) component and the spectrum of baryon acoustic oscillations. Additional radiation alters the sound horizon, which is routinely captured in current BAO~analyses. While this is highly degenerate with other parameters, combining BAO~and CMB~observations slightly improves the sensitivity to~$\Neff$ over the~CMB alone, $\Neff = 2.99 \pm 0.17$~\cite{Aghanim:2018eyx}. The BAO~spectrum also exhibits the same phase shift observed in the CMB~spectra. A nonzero phase shift was recently extracted from the distribution of galaxies observed by the Baryon Oscillation Spectroscopic Survey~(BOSS)~\cite{Baumann:2017gkg, Baumann:2018qnt} and upcoming galaxy surveys will significantly improve on this measurement.

The two main consequences of a different radiation density on the broadband shape of the power spectrum are a change of the power on small scales and in the location of the turn-over of the spectrum. Although these effects are clearly visible in the linear matter power spectrum, they are limited by uncertainties related to gravitational nonlinearities and biasing. The combination of planned spectroscopic LSS~surveys with \planck\ data could reach $\sigma(\Neff)=0.08$~\cite{Baumann:2017gkg}. However, these surveys would not contribute a meaningful improvement when combined with a CMB~experiment achieving $\sigma(\Neff) \approx 0.03$. If nonlinear effects can be controlled, very large-volume and high-resolution LSS~maps can reach comparable sensitivity to the CMB~\cite{Baumann:2017gkg, Ansari:2018ury} and would significantly add to the scientific impact of the~CMB alone. Furthermore, LSS~observations are also sensitive to effects induced by neutrinos and other light relics beyond~$\Neff$, for example in the Lyman-$\alpha$ forest and the biasing of galaxies %
(see e.g.~\cite{Boyarsky:2008xj, Zhu:2014qma, LoVerde:2014pxa, Munoz:2018ajr, Chiang:2018laa}).

\vskip4pt
\noindent\textbf{Summary}\hskip8pt
\emph{Sub-percent-level measurements of the radiation density would transform our understanding of the early universe, the neutrino and dark sectors, and more. To reach clear observational targets, future CMB~observations offer the most promising and concrete path in the next decade.}

\clearpage
\section*{Affiliations}
\newcommand{\affiliation}{\footnote{}}

\newcommand{\Amherst}{University of Massachusetts, Amherst, MA 01003 USA}
\newcommand{\ANLHEP}{HEP Division, Argonne National Laboratory, Lemont, IL~60439, USA}
\newcommand{\APC}{Laboratoire Astroparticule et Cosmologie (APC), CNRS/IN2P3, Universit\'e Paris Diderot, 75205~Paris, France}
\newcommand{\ASU}{Arizona State University, Tempe, AZ  85287}
\newcommand{\BenGurion}{Department of Physics, Ben-Gurion University, Be'er~Sheva~84105, Israel}
\newcommand{\BNL}{Brookhaven National Laboratory, Upton, NY~11973, USA}
\newcommand{\Brown}{Brown University, Providence, RI~02912, USA}
\newcommand{\Bub}{Boston University, Boston, MA 02215}
\newcommand{\BU}{Boston University, Boston, MA~02215, USA}
\newcommand{\Buffalo}{Department of Physics, University at Buffalo, SUNY Buffalo, NY 14260 USA}
\newcommand{\Caltech}{California Institute of Technology, Pasadena, CA~91125, USA}
\newcommand{\Cardiff}{School of Physics and Astronomy, Cardiff University, The Parade, Cardiff, CF24 3AA, UK}
\newcommand{\Carleton}{Carleton University, K1S 5B6 Ottawa, Canada}
\newcommand{\Carnegie}{The Observatories of the Carnegie Institution for Science, 813 Santa Barbara St., Pasadena, CA 91101, USA}
\newcommand{\Cavendish}{Astrophysics Group, Cavendish Laboratory, University of Cambridge, Cambridge~CB3~0HE, UK}
\newcommand{\CCA}{Center for Computational Astrophysics, Flatiron Institute, New York, NY~10010, USA}
\newcommand{\CPPM}{CPPM, Aix-Marseille Universit\'e, CNRS/IN2P3, 13007~Marseille, France}
\newcommand{\CEADAP}{D\'epartement d’Astrophysique, CEA Saclay DSM/Irfu, 91191 Gif-sur-Yvette, France}
\newcommand{\CERN}{CERN, Geneva, Switzerland}
\newcommand{\CfA}{Harvard-Smithsonian Center for Astrophysics, Cambridge, MA~02138, USA}
\newcommand{\CFT}{Center for Theoretical Physics, Polish Academy of Sciences, al. Lotnik\'{o}w 32/46, 02-668, Warsaw, Poland}
\newcommand{\Cincinnati}{University of Cincinnati, Cincinnati, OH~45221, USA}
\newcommand{\CITA}{Canadian Institute for Theoretical Astrophysics, University of Toronto, Toronto, ON~M5S~3H8, Canada}
\newcommand{\CNRSA}{CNRS, Laboratoire d'Annecy-le-Vieux de Physique Th\'{e}orique, France}
\newcommand{\CNYang}{C.N. Yang Institute for Theoretical Physics State University of New York Stony Brook, NY 11794}
\newcommand{\CMUCosmo}{Department of Physics, McWilliams Center for Cosmology, Carnegie Mellon University}
\newcommand{\Columbia}{Columbia University, New York, NY~10027, USA}
\newcommand{\Cornell}{Cornell University, Ithaca, NY~14853, USA}
\newcommand{\CPthree}{CP3-Origins, 5230 Odense, Denmark}
\newcommand{\CWRU}{Case Western Reserve University, Cleveland, OH~44106, USA}
\newcommand{\daa}{Department of Astronomy and Astrophysics, University of Toronto, ON~M5S~3H4, Canada}
\newcommand{\damtp}{DAMTP, University of Cambridge, Cambridge CB3~0WA, UK}
\newcommand{\DESY}{DESY,  22607 Hamburg, Germany}
\newcommand{\DFI}{Departamento de F\'isica, FCFM, Universidad de Chile, Santiago, Chile}
\newcommand{\DOE}{US. Department of Energy, Germantown, MD 20874}
\newcommand{\drexel}{Drexel University, Philadelphia, PA 19104}
\newcommand{\Duke}{Duke University and Triangle Universitites Nuclear Laboratory, Durham, NC 27708}
\newcommand{\DukePhys}{Department of Physics, Duke University, Durham, NC 27708, USA}
\newcommand{\dunlap}{Dunlap Institute for Astronomy and Astrophysics, University of Toronto, ON~M5S~3H4, Canada}
\newcommand{\Durham}{Department of Physics, Durham University, Durham DH1~3LE, UK}
\newcommand{\ED}{University of Edinburgh, Edinburgh EH8~9YL, UK}
\newcommand{\EPFL}{Institute of Physics, Laboratory of Astrophysics, École Polytechnique Fédérale de Lausanne (EPFL), Observatoire de Sauverny, 1290~Versoix, Switzerland}
\newcommand{\ETH}{ETH Zurich, Institute for Particle Physics, 8093 Zurich, Switzerland}
\newcommand{\FNAL}{Fermi National Accelerator Laboratory, Batavia, IL~60510, USA}
\newcommand{\FQAUB}{Dept. de F\' isica Qu\` antica i Astrof\' isica, Universitat de Barcelona, Mart\' i i Franqu\` es 1, E08028 Barcelona, Spain}
\newcommand{\FSU}{Florida State University, Tallahassee, FL 32306}
\newcommand{\Glasgow}{University of Glasgow, G12 8QQ Glasgow, United Kingdom}
\newcommand{\GRAPPA}{GRAPPA, University of Amsterdam, 1098~XH~Amsterdam, The Netherlands}
\newcommand{\GSFC}{Goddard Space Flight Center, Greenbelt, MD~20771, USA}
\newcommand{\GWU}{George Washington University, Washington, DC 20052}
\newcommand{\Hampton}{Hampton University, Hampton, VA 23668}
\newcommand{\HarvardPhys}{Department of Physics, Harvard University, Cambridge, MA~02138, USA}
\newcommand{\Haverford}{Haverford College, Haverford, PA~19041, USA}
\newcommand{\Hawaii}{University of Hawaii, Honolulu, HI 96822}
\newcommand{\HKUST}{The Hong Kong University of Science and Technology, Hong~Kong~SAR, China}
\newcommand{\houston}{University of Houston, Houston, TX 77204}
\newcommand{\IAP}{Institut d'Astrophysique de Paris (IAP), CNRS \& Sorbonne University, 75014~Paris, France}
\newcommand{\IAS}{Institute for Advanced Study, Princeton, NJ~08540, USA}
\newcommand{\IBS}{Institute for Basic Science (IBS), Daejeon~34051, Korea}
\newcommand{\ICC}{ICC, University of Barcelona, IEEC-UB, Mart\' i i Franqu\` es, 1, E08028 Barcelona, Spain}
\newcommand{\ICCD}{Institute for Computational Cosmology, Department of Physics, Durham University, South Road, Durham, DH1 3LE, UK}
\newcommand{\ICE}{Institute of Space Sciences (ICE, CSIC), Campus UAB, Carrer de Can Magrans, s/n, 08193 Barcelona, Spain}
\newcommand{\ICRR}{Institute for Cosmic Ray Resaerch, The University of Tokyo, 456 Higashi-Mozumi, Kamioka, Hida, Gifu 506-1205, Japan}
\newcommand{\ICTP}{International Centre for Theoretical Physics (ICTP), 34151~Trieste, Italy}
\newcommand{\IFAE}{Institut de Fisica d’Altes Energies, The Barcelona Institute of Science and Technology, Campus UAB, 08193 Bellaterra (Barcelona), Spain}
\newcommand{\IFPU}{Institute for Fundamental Physics of the Universe (IFPU), 34014~Trieste, Italy}
\newcommand{\IFT}{Instituto de Fisica Teorica UAM/CSIC, Universidad Autonoma de Madrid, 28049 Madrid, Spain}
\newcommand{\IFUNAM}{Instituto de F\'{i}sica (IFUNAM), Universidad Nacional Aut\'onoma de M\'exico, 04510~Ciudad de México, Mexico}
\newcommand{\IHEP}{Institute of High Energy Physics, Austrian Academy of Sciences, 1050 Vienna, Austria}
\newcommand{\Imperial}{Theoretical Physics, Blackett Laboratory, Imperial College, London SW7~2AZ, UK}
\newcommand{\Indiana}{Indiana University, Bloomington, IN 47405}
\newcommand{\INAFOATs}{INAF - Osservatorio Astronomico di Trieste, Via G.B. Tiepolo 11, 34143 Trieste, Italy}
\newcommand{\INAFOAS}{INAF - Osservatorio di Astrofisica e Scienza dello Spazio di Bologna, via Piero Gobetti 93/3, I-40129 Bologna, Italy}
\newcommand{\INFNCag}{Istituto Nazionale di Fisica Nucleare, Sezione di Cagliari,  09126 Cagliari, Italy}
\newcommand{\INFNCat}{Istituto Nazionale di Fisica Nucleare, Sezione di Catania, 95125 Catania, Italy}
\newcommand{\INFNG}{Istituto Nazionale di Fisica Nucleare, Sezione di Genova, 16146 Genova, Italy}
\newcommand{\INFN}{National Institute for Nuclear Physics (INFN), 34127~Trieste, Italy}
\newcommand{\INFNFE}{Istituto Nazionale di Fisica Nucleare, Sezione di Ferrara, 40122~Ferrara, Italy}
\newcommand{\INFNLNF}{Istituto Nazionale di Fisica Nucleare, Laboratori Nazionali di Frascati, 00044 Frascati, Italy}
\newcommand{\INFNLNS}{Istituto Nazionale di Fisica Nucleare, Laboratori Nazionali del Sud, 95125 Catania, Italy}
\newcommand{\INFNN}{Istituto Nazionale di Fisica Nucleare, Sezione di Napoli, 80125 Napoli, Italy }
\newcommand{\INFNRM}{Istituto Nazionale di Fisica Nucleare, Sezione di Roma, 00185~Roma, Italy}
\newcommand{\INFNT}{Istituto Nazionale di Fisica Nucleare, Sezione di Torino, 10125, Italy }
\newcommand{\ioa}{Institute of Astronomy, University of Cambridge, Cambridge CB3~0HA, UK}
\newcommand{\IPP}{Institute for Particle Physics, BC V8W 3P6 Victoria, Canada}
\newcommand{\IPMU}{Kavli Institute for the Physics and Mathematics of the Universe (WPI), University of Tokyo, 277-8583~Kashiwa, Japan}
\newcommand{\IPNL}{Universit\'e de Lyon, F-69622, Lyon, France; Universit\'e de Lyon 1, Villeurbanne; CNRS/IN2P3, Institut de Physique Nucl\'eaire de Lyon}
\newcommand{\IRFU}{IRFU, CEA, Universit\'e Paris-Saclay, F-91191 Gif-sur-Yvette, France}
\newcommand{\ITFA}{Institute for Theoretical Physics, University of Amsterdam, 1098~XH~Amsterdam, The Netherlands}
\newcommand{\IUCAA}{The Inter-University Centre for Astronomy and Astrophysics, Pune, 411007, India}
\newcommand{\Jerusalem}{Hebrew University of Jerusalem, 91904 Jerusalem, Israel}
\newcommand{\JHU}{Johns Hopkins University, Baltimore, MD~21218, USA}
\newcommand{\JLAB}{Thomas Jefferson National Laboratory, Newport News, VA 23606}
\newcommand{\JPL}{Jet Propulsion Laboratory, California Institute of Technology, Pasadena, CA~91109, USA}
\newcommand{\KASSI}{Korea Astronomy and Space Science Institute, Daejeon~34055, Korea}
\newcommand{\kavli}{Kavli Institute for Cosmology, Cambridge CB3~0HA, UK}
\newcommand{\KIAS}{School of Physics, Korea Institute for Advanced Study, 85 Hoegiro, Dongdaemun-gu, Seoul 130-722, Korea}
\newcommand{\KICP}{Kavli Institute for Cosmological Physics, Chicago, IL~60637, USA}
\newcommand{\KIPAC}{Kavli Institute for Particle Astrophysics and Cosmology, Stanford, CA~94305, USA}
\newcommand{\KINGS}{King's College London, WC2R 2LS London, United Kingdom}
\newcommand{\Kobe}{Kobe University, 657-8501 Kobe, Japan}
\newcommand{\KPH}{Johannes Gutenberg University, 55128 Mainz, Germany}
\newcommand{\KPMU}{University of Tokyo, 277-8583  Kashiwa , Japan}
\newcommand{\KSU}{Kansas State University, Manhattan, KS~66506, USA}
\newcommand{\Lafayette}{Lafayette College, Easton, PA 18042}
\newcommand{\LANL}{Los Alamos National Laboratory, Los Alamos, NM 87545}
\newcommand{\LBL}{Lawrence Berkeley National Laboratory, Berkeley, CA~94720, USA}
\newcommand{\Leiden}{Lorentz Institute, Leiden University, Niels Bohrweg 2,Leiden, NL 2333 CA, The Netherlands}
\newcommand{\Liverpool}{University of Liverpool,  L69 7ZE Liverpool , United Kingdom}
\newcommand{\LLNL}{Lawrence Livermore National Laboratory, Livermore, CA~94550, USA}
\newcommand{\LPC}{Universit\'e Clermont Auvergne, CNRS/IN2P3, Laboratoire de Physique de Clermont, F-63000 Clermont-Ferrand, France}
\newcommand{\LPNHE}{Sorbonne Universit\'e, Universit\'e Paris Diderot, CNRS/IN2P3, Laboratoire de Physique Nucl\'eaire et de Hautes Energies, LPNHE, 4 Place Jussieu, F-75252 Paris, France}
\newcommand{\McGill}{McGill University, Montreal, QC H3A 2T8, Canada}
\newcommand{\Melbourne}{School of Physics, The University of Melbourne, Parkville, VIC~3010, Australia}
\newcommand{\Mines}{Colorado School of Mines, Golden, CO 80401}
\newcommand{\MIT}{Massachusetts Institute of Technology, Cambridge, MA~02139, USA}
\newcommand{\MPE}{Max-Planck-Institut f\"{u}r extraterrestrische Physik (MPE), Giessenbachstrasse 1, D-85748 Garching bei M\"unchen, Germany}
\newcommand{\MPIA}{Max-Planck-Institut f\"{u}r Astrophysik, Karl-Schwarzschild-Str. 1, 85741 Garching, Germany}
\newcommand{\MPP}{Max-Planck-Institut f\"{u}r Physik (Werner-Heisenberg-Institut), F\"ohringer Ring 6, D-80805 M\"unchen, Germany}
\newcommand{\LUPM}{Laboratoire Univers et Particules de Montpellier, Univ. Montpellier and CNRS, 34090 Montpellier, France}
\newcommand{\NAOC}{National Astronomical Observatories, Chinese Academy of Sciences, Beijing, China}
\newcommand{\NCBJ}{National Center for Nuclear Research, 02-093~Warsaw, Poland}
\newcommand{\NCU}{National Central University, Taoyuan City 32001, Taiwan (R.O.C.)}
\newcommand{\NCSU}{Physics Department, North Carolina State Universitym, 2401 Stinson Dr, Raleigh, NC 27607}
\newcommand{\ND}{University of Notre Dame,vNotre Dame, IN 46556}
\newcommand{\NIU}{Northern Illinois University, DeKalb, Illinois 60115}
\newcommand{\NMSU}{New Mexico State University, Las Cruces, NM 88003}
\newcommand{\NOAO}{National Optical Astronomy Observatory, 950 N. Cherry Ave., Tucson, AZ 85719 USA}
\newcommand{\Northwestern}{Northwestern University, Evanston, IL 60201}
\newcommand{\Nottingham}{University of Nottingham, NG7 2RD Nottingham, United Kingdom}
\newcommand{\NWU}{Northwestern University, Evanston, IL 60208}
\newcommand{\NYU}{New York University, New York, NY 10003}
\newcommand{\OK}{ University of Oklahoma, Norman, OK 73019}
\newcommand{\ORNL}{Oak Ridge National Laboratory, Oak Ridge, TN 37831}
\newcommand{\OSU}{The Ohio State University, Columbus, OH~43212, USA}
\newcommand{\OU}{Department of Physics and Astronomy, Ohio University, Athens, OH~45701, USA}
\newcommand{\OskarKlein}{Oskar Klein Centre for Cosmoparticle Physics, Stockholm University, AlbaNova, 106~91~Stockholm, Sweden}
\newcommand{\Oxford}{University of Oxford, Oxford OX1~3RH, UK}
\newcommand{\Oxy}{Occidental College, Los Angeles, CA 90041}
\newcommand{\ParisSud}{LAL, Universit\'{e} Paris-Sud, 91898~Orsay~Cedex, France \& CNRS/IN2P3, 91405~Orsay, France}
\newcommand{\PI}{Perimeter Institute, Waterloo, ON~N2L~2Y5, Canada}
\newcommand{\Pitt}{University of Pittsburgh and PITT PACC, Pittsburgh, PA 15260}
\newcommand{\PNNL}{Pacific Northwest National Laboratory ,Richland, WA 99352}
\newcommand{\PNPI}{Petersburg Nuclear Physics Institute, 188300 Gatchina, Russia}
\newcommand{\Port}{Institute of Cosmology \& Gravitation, University of Portsmouth, Portsmouth PO1~3FX, UK}
\newcommand{\Princeton}{Princeton University, Princeton, NJ~08544, USA}
\newcommand{\PSU}{The Pennsylvania State University, University Park, PA 16802}
\newcommand{\Purdue}{Purdue University, West Lafayette, IN 47907}
\newcommand{\PW}{Participation Worldscope, Sedona, Arizona and Hong Kong, SAR PRC}
\newcommand{\Queens}{Queen's University , K7L 3N6 Kingston, Canada}
\newcommand{\Queensland}{The University of Queensland, School of Mathematics and Physics, QLD 4072, Australia}
\newcommand{\QMUL}{Queen Mary University of London, London E1~4NS, UK}
\newcommand{\RAL}{Radio Astronomy Laboratory, University of California Berkeley, Berkeley, CA 94720, USA}
\newcommand{\Rice}{Department of Physics \& Astronomy, Rice University, Houston, TX~77005, USA}
\newcommand{\RIT}{Rochester Institute of Technology}
\newcommand{\RomaS}{Dipartimento di Fisica, Universit\`{a} La Sapienza, 00185~Roma, Italy}
\newcommand{\RUG}{Kapteyn Astronomical Institute, University of Groningen, P.O. Box 800, 9700 AV Groningen, The Netherlands}
\newcommand{\Rutgers}{Department of Physics and Astronomy, Rutgers, the State University of New Jersey, 136 Frelinghuysen Road, Piscataway, NJ 08854, USA}
\newcommand{\Sanford}{Sanford Underground Research Facility, Lead, SD 57754}
\newcommand{\Sassari}{Universit\`a di Sassari, 07100 Sassari,  Italy}
\newcommand{\SCIPP}{University of California at Santa Cruz, Santa Cruz, CA 95064}
\newcommand{\Sejong}{Department of Physics and Astronomy, Sejong University, Seoul~143-747, Korea}
\newcommand{\Sheffield}{University of Sheffield, S3 7RH Sheffield, United Kingdom}
\newcommand{\SHAO}{Shanghai Astronomical Observatory (SHAO), Shanghai~200030, China}
\newcommand{\Siena}{Siena College, 515 Loudon Road, Loudonville, NY 12211, USA}
\newcommand{\SISSA}{International School for Advanced Studies~(SISSA), 34136~Trieste, Italy}
\newcommand{\SimonFraser}{Department of Physics, Simon Fraser University, Burnaby, BC~V5A~1S6, Canada}
\newcommand{\SLAC}{SLAC National Accelerator Laboratory, Menlo Park, CA~94025, USA}
\newcommand{\SMU}{Southern Methodist University, Dallas, TX~75275, USA}
\newcommand{\SNOLAB}{SNOLAB, Lively, ON P3Y 1N2, Canada}
\newcommand{\SoCal}{University of Southern California, Los Angeles, CA~90089, USA}
\newcommand{\Stanford}{Stanford University, Stanford, CA~94305, USA}
\newcommand{\StonyBrook}{Stony Brook University, Stony Brook, NY~11794, USA}
\newcommand{\STSCI}{Space Telescope Science Institute, Baltimore, MD~21218, USA}
\newcommand{\SUNYA}{University at Albany SUNY, Albany, NY 12222}
\newcommand{\SussexAstronomy}{Astronomy Centre, School of Mathematical and Physical Sciences, University of Sussex, Brighton BN1~9QH, UK}
\newcommand{\Syracuse}{Syracuse University, Syracuse, NY~13244, USA}
\newcommand{\Tamu}{Texas AandM University, College Station, TX 77843 }
\newcommand{\Techsource}{Techsource Incorporated, Los Alamos, NM 87544}
\newcommand{\TelAviv}{Tel-Aviv University,  69978 Tel-Aviv, Israel}
\newcommand{\Temple}{Temple University, Philadelphia, PA 19122}
\newcommand{\TIFR}{Tata Institute of Fundamental Research, Mumbai~400005, India}
\newcommand{\Tsinghua}{Department of Physics and Tsinghua Center for Astrophysics, Tsinghua University, Beijing 100084, P R China}
\newcommand{\TUM}{Technical University of Munich,  80333 Munich, Germany}
\newcommand{\UA}{University of Alabama, Tuscaloosa, AL 35487}
\newcommand{\UAS}{Department of Astronomy/Steward Observatory, University of Arizona, Tucson, AZ  85721}
\newcommand{\UAM}{Universidad Aut\'onoma de Madrid, 28049~Madrid, Spain}
\newcommand{\UBC}{University of British Columbia, Vancouver, BC V6T 1Z1, Canada}
\newcommand{\UCB}{Department of Astronomy, University of California Berkeley, Berkeley, CA~94720, USA}
\newcommand{\UCBP}{Department of Physics, University of California Berkeley, Berkeley, CA~94720, USA}
\newcommand{\UCBSSL}{Space Sciences Laboratory, University of California Berkeley, Berkeley, CA~94720, USA}
\newcommand{\UCD}{University of California Davis, Davis, CA~95616, USA}
\newcommand{\UChicago}{University of Chicago, Chicago, IL~60637, USA}
\newcommand{\UCI}{University of California Irvine, Irvine, CA~92697, USA}
\newcommand{\UCLA}{University of California Los Angeles, Los Angeles, CA~90095, USA}
\newcommand{\UCL}{University College London, London WC1E~6BT, UK}
\newcommand{\UCR}{University of California Riverside, Riverside, CA~92521, USA}
\newcommand{\UCSB}{University of California Santa Barbara, Santa Barbara, CA~93106, USA}
\newcommand{\UCSC}{University of California Santa Cruz, Santa Cruz, CA 95064, USA}
\newcommand{\UCSD}{University of California San Diego, La Jolla, CA~92093, USA}
\newcommand{\UFL}{University of Florida, Gainesville, FL~32611, USA}
\newcommand{\UFN}{Universit\`a Federico II di Napoli, 80125 Napoli, Italy}
\newcommand{\UGTO}{Divisi\'on de Ciencias e Ingenier\'ias, Universidad de Guanajuato, Le\'on~37150, M\'exico}
\newcommand{\UKY}{University of Kentucky, Lexington, KY 40506}
\newcommand{\UMD}{University of Maryland, College Park, MD 20742
	\newcommand{\UMiami}{University of Miami, Coral Gables, FL 33124}}
\newcommand{\UMich}{University of Michigan, Ann Arbor, MI~48109, USA}
\newcommand{\UMN}{University of Minnesota, Minneapolis, MN 55455, USA}
\newcommand{\UnB}{Instituto de F\'{i}sica, Universidade de Bras\'{i}lia, 70919-970~Bras\'{i}lia, Brazil}
\newcommand{\UNC}{University of North Carolina at Chapel Hill, Chapel Hill, NC 27599}
\newcommand{\UNH}{University of New Hampshire, Durham, NH 03824}
\newcommand{\UNIMI}{Dipartimento di Fisica ``Aldo Pontremoli'', Universit\`a{} degli Studi di Milano, via Celoria 16, 20133 Milano, Italy}
\newcommand{\UNIPD}{Dipartimento di Fisica e Astronomia ``G. Galilei'', Universit\`a degli Studi di Padova, 35131~Padova, Italy}
\newcommand{\UNM}{University of New Mexico, Albuquerque, NM~87131, USA}
\newcommand{\UNV}{University of Nevada, Reno, NV 89557}
\newcommand{\UoM}{Jodrell Bank Center for Astrophysics, School of Physics and Astronomy, University of Manchester, Manchester M13~9PL, UK}
\newcommand{\UPenn}{Department of Physics and Astronomy, University of Pennsylvania, Philadelphia, PA~19104, USA}
\newcommand{\UR}{Department of Physics and Astronomy, University of Rochester, Rochester, NY~14627, USA}
\newcommand{\UrbanaC}{Department of Physics, University of Illinois at Urbana-Champaign, Urbana, IL~61801, USA}
\newcommand{\USC}{The University of South Carolina, Columbia, SC 29208}
\newcommand{\USD}{The University of South Dakota, Vermillion, SD 57069}
\newcommand{\UTD}{University of Texas at Dallas, Richardson, TX~75080, USA}
\newcommand{\Utenn}{The University of Tennessee, Knoxville, TN 37996}
\newcommand{\Utah}{University of Utah, Department of Physics and Astronomy, 115 S 1400 E, Salt Lake City, UT 84112, USA}
\newcommand{\UVA}{University of Virginia, Charlottesville, VA 22903}
\newcommand{\Uvic}{University of Victoria, BC V8P 5C2 Victoria, Canada}
\newcommand{\UWaterloo}{Department of Physics and Astronomy, University of Waterloo, Waterloo, ON~N2L~3G1, Canada}
\newcommand{\UWMadison}{Department of Physics, University of Wisconsin-Madison, Madison, WI~53706, USA}
\newcommand{\UW}{University of Washington, Seattle, WA~98195, USA}
\newcommand{\UWC}{Department of Physics \& Astronomy, University of the Western Cape, Cape Town 7535, South Africa}
\newcommand{\Vanderbilt}{Physics \& Astronomy Department, Vanderbilt University, Nashville, TN~37235, USA}
\newcommand{\VSI}{Van Swinderen Institute for Particle Physics and Gravity, University of Groningen, 9747~AG~Groningen, The~Netherlands}
\newcommand{\VT}{Virginia Tech, Blacksburg, VA~24061, USA}
\newcommand{\VUU}{Virginia Union University, Richmond, Virginia, 23220}
\newcommand{\WCA}{Centre for Astrophysics, University of Waterloo, Waterloo, ON~N2L~3G1, Canada}
\newcommand{\Weizmann}{Weizmann Institute of Science, 76100 Rehovot, Israel}
\newcommand{\Wellesley}{Wellesley College, Wellesley, MA 02481}
\newcommand{\wiscIce}{University of Wisconsin, Madison, WI 53706}
\newcommand{\WM}{College of William and Mary, Newport News, VA 23606}
\newcommand{\WUSL}{Washington University in St Louis, St. Louis, MO 63130}
\newcommand{\WVU}{CSEE, West Virginia University, Morgantown, WV~26505, USA}
\newcommand{\WVUGWAC}{Center for Gravitational Waves and Cosmology, West Virginia University, Morgantown, WV~26505, USA}
\newcommand{\Wyoming}{Department of Physics and Astronomy, University of Wyoming, Laramie, WY~82071, USA}
\newcommand{\Yale}{Department of Physics, Yale University, New Haven, CT~06520, USA}
\newcommand{\YorkU}{Department of Physics and Astronomy, York University, Toronto, Ontario M3J 1P3, Canada}
\newcommand{\Amsterdam}{Department of Physics, Science Park, University of Amsterdam - the Netherlands}
\newcommand{\Stockholm}{The Oskar Klein Centre for Cosmoparticle Physics,
	Department of Physics, Stockholm University, SE-106 91 Stockholm, Sweden}
\newcommand{\Stonybrook}{Stonybrook}
\newcommand{\lancaster}{Consortium for Fundamental Physics, Physics Department, Lancaster University, Lancaster LA1 4YB, UK}
\newcommand{\Unige}{Department of Theoretical Physics and Center for Astroparticle
	Physics, University of Geneva, 24 quai E. Ansermet, CH-1211 Geneva 4, Switzerland}
\newcommand{\IAPdeux}{Institut Lagrange de Paris, Sorbonne Universit ́es, 98 bis Boulevard Arago, 75014 Paris, France}
\newcommand{\Sussex}{Department of Physics & Astronomy, University of Sussex, Brighton BN1 9QH, United Kingdom}
\newcommand{\Chile}{Grupo de Cosmolog ́ıa y Astrof ́ısica Teo ́rica, Departamento de F ́ısica,
	FCFM, Universidad de Chile, Blanco Encalada 2008, Santiago, Chile}
\newcommand{\SEJONG}{Department of Physics and Astronomy, Sejong University, Seoul, 143-747}
\newcommand{\NBI}{The Niels Bohr Institute & Discovery Center, Blegdamsvej 17, DK-2100 Copenhagen, Denmark}

\begin{raggedright}
\setlength{\columnsep}{18pt}
\begin{multicols}{2}
\footnotesize

\noindent
$^{1}$ \Rice \\
$^{2}$ \SMU \\
$^{3}$ \IAS \\
$^{4}$ \UCSD \\
$^{5}$ \UCI \\
$^{6}$ \damtp \\
$^{7}$ \UrbanaC \\
$^{8}$ \SLAC \\
$^{9}$ \Durham \\
$^{10}$ \LLNL \\
$^{11}$ \SISSA \\
$^{12}$ \IFPU \\
$^{13}$ \INFN \\
$^{14}$ \WVU \\
$^{15}$ \WVUGWAC \\
$^{16}$ \UNM \\
$^{17}$ \Cornell \\
$^{18}$ \GRAPPA \\
$^{19}$ \ITFA \\
$^{20}$ \UWMadison \\
$^{21}$ \JHU \\
$^{22}$ \FNAL \\
$^{23}$ \KICP \\
$^{24}$ \Port \\
$^{25}$ \Cincinnati \\
$^{26}$ \ANLHEP \\
$^{27}$ \CITA \\
$^{28}$ \LBL \\
$^{29}$ \PI \\
$^{30}$ \UChicago \\
$^{31}$ \UCBP \\
$^{32}$ \ioa \\
$^{33}$ \kavli \\
$^{34}$ \CfA \\
$^{35}$ \UCSB \\
$^{36}$ \HarvardPhys \\
$^{37}$ \Stanford \\
$^{38}$ \JPL \\
$^{39}$ \BU \\
$^{40}$ \Princeton \\
$^{41}$ \UCR \\
$^{42}$ \CPPM \\
$^{43}$ \GSFC \\
$^{44}$ \SussexAstronomy \\
$^{45}$ \UAM \\
$^{46}$ \UFL \\
$^{47}$ \UR \\
$^{48}$ \Haverford \\
$^{49}$ \OskarKlein \\
$^{50}$ \UMN \\
$^{51}$ \Cavendish \\
$^{52}$ \CCA \\
$^{53}$ \OSU \\
$^{54}$ \dunlap \\
$^{55}$ \daa \\
$^{56}$ \VT \\
$^{57}$ \UMich \\
$^{58}$ \IBS \\
$^{59}$ \KASSI \\
$^{60}$ \TIFR \\
$^{61}$ \EPFL \\
$^{62}$ \UCD \\
$^{63}$ \Brown \\
$^{64}$ \BenGurion \\
$^{65}$ \UCL \\
$^{66}$ \INFNFE \\
$^{67}$ \UNIPD \\
$^{68}$ \StonyBrook \\
$^{69}$ \MIT \\
$^{70}$ \UW \\
$^{71}$ \VSI \\
$^{72}$ \ICTP \\
$^{73}$ \IAP \\
$^{74}$ \Yale \\
$^{75}$ \BNL \\
$^{76}$ \RomaS \\
$^{77}$ \INFNRM \\
$^{78}$ \SoCal \\
$^{79}$ \SimonFraser \\
$^{80}$ \KIPAC \\
$^{81}$ \APC \\
$^{82}$ \Melbourne \\
$^{83}$ \STSCI \\
$^{84}$ \Sejong \\
$^{85}$ \CWRU \\
$^{86}$ \UCB \\
$^{87}$ \OU \\
$^{88}$ \ParisSud \\
$^{89}$ \UPenn \\
$^{90}$ \UoM \\
$^{91}$ \IFUNAM \\
$^{92}$ \Syracuse \\
$^{93}$ \Vanderbilt \\
$^{94}$ \UCLA \\
$^{95}$ \NAOC \\
$^{96}$ \ED  \\
\normalsize
\end{multicols}
\end{raggedright}

\newpage
\makeatletter
\renewcommand\section{\@startsection{section}{1}{\z@}%
	{-3.5ex \@plus -1.3ex \@minus -.7ex}%
	{2.3ex \@plus.4ex \@minus .4ex}%
	{\normalfont\large\bfseries}}
\makeatother
\bibliographystyle{utphys}
\bibliography{references}
\end{document}